\documentclass[amsmath,amssymb,twocolumn,amsfonts,showpacs]{revtex4}
\usepackage{graphicx}
\def \beq {\begin{equation}}
\def \eeq {\end{equation}}

\newcommand{\ga}{\gamma_{se}^{12}}
\newcommand{\gb}{\gamma_{se}^{21}}
\newcommand{\EE}{\mathbb{E}}

\begin{document}
\title{Spin-noise correlations and spin-noise exchange driven by low-field spin-exchange collisions}
\author{A. T. Dellis$^1$, M. Loulakis$^2$ and I. K. Kominis$^1$}
\email{ikominis@physics.uoc.gr}
\affiliation{$^1$Department of Physics, University of Crete, 71103 Heraklion, Greece\\
$^2$School of Applied Mathematical and Physical Sciences, National Technical University of Athens, 15780 Athens, Greece}

\begin{abstract}
The physics of spin exchange collisions have fueled several discoveries in fundamental physics and numerous applications in medical imaging and nuclear magnetic resonance. We here report on the experimental observation and theoretical justification of spin-noise exchange, the transfer of spin-noise from one atomic species to another. The signature of spin-noise exchange is an increase of the total spin-noise power at low magnetic fields, on the order of 1 mG,  where the two-species spin-noise resonances overlap. The underlying physical mechanism is the two-species spin-noise correlation induced by spin-exchange collisions.
\end{abstract}
\maketitle 
\section{Introduction}
The Pauli exchange interaction, of fundamental importance for understanding the structure of matter, also underlies spin-dependent atomic collisions \cite{happer,happer_walker}. Spin-exchange collisions in atomic vapors  have fueled a wide range of scientific investigations, ranging from enhanced NMR signals and new MRI techniques \cite{cates,albert,navon} to nuclear scattering experiments sensitive to the nuclear or nucleon spin structure \cite{accel}. Many of the aforementioned phenomena rely on the spin-exchange transfer of large spin polarizations from one atomic species to another. 

We here extend spin exchange into a deeper layer of collective spin degrees of freedom, namely we demonstrate the transfer of quantum spin fluctuations from one atomic species to another, a phenomenon we term spin-noise exchange. Quantum fluctuations  and their interspecies transfer are central to emerging technologies of quantum information, like quantum memories using atomic spin or pseudo-spin ensembles \cite{memory1,memory2}. Spin noise \cite{sn1}, in particular,  determines the quantum limits to the precision of atomic vapor clocks \cite{clocks} and the sensitivity of atomic magnetometers \cite{magn1,magn2,kitching,budker_romalis}, the most recent of which utilize several spin species \cite{romalis_comagn}. The fundamental understanding of spin-noise exchange could have further repercussions, from noise-energy harvesting in spintronic devices \cite{gammaitoni}, to novel spin-dependent phenomena in intergalactic hydrogen gas \cite{icm}. A similar effect to the one described herein was observed with solid-state nuclear spins \cite{poggio1,poggio2}, but the transfer of nuclear spin fluctuations was evoked with externally applied magnetic fields. In our case the transfer is spontaneous and driven by incessant atomic spin-exchange collisions.

Spin-exchange collisions are central to optical pumping of atomic vapors \cite{happer_book}. Even without externally manipulating atoms with light, i.e. leaving them in an unpolarized equilibrium state, spin-exchange collisions lead to continuous spin fluctuations around the average value of zero. Such spontaneous spin noise has been recently demonstrated \cite{crooker,oestreich,crooker2,sherman,zapasskii,roy,li} to be a versatile spectroscopic tool in atomic and condensed matter physics. In particular, spin noise in a rubidium vapor was measured \cite{crooker} at a magnetic field of several Gauss, allowing the spin-noise resonances of $^{85}$Rb and $^{87}$Rb (occurring approximately in the ratio 3:1 in rubidium of natural abundance) to be clearly distinguished. This is so since the respective gyromagnetic ratios are $g_{1}=466~{\rm kHz/G}$ and $g_{2}=700~{\rm kHz/G}$, whereas the resonance line width was on the order of 10 kHz. 

The total area under the spectral distribution of spin-noise power is the total spin variance, intuitively expected to be constant, i.e. independent of the magnetic field at which the measurement is performed, or equivalently, independent of where along the frequency axis the two spin resonances are positioned. 

We will here demonstrate experimentally and prove theoretically that the total spin-noise power of a two-species spin ensemble, like $^{85}$Rb -$^{87}$Rb, exhibits a counter-intuitive dependence on the applied magnetic field. This is the experimental signature of spin-noise exchange, which is observable when the two atomic species have overlapping spin-noise resonances. For the resonance width in our measurement, of about 1 kHz, this overlap happens at magnetic fields on the order of 1 mG. 

In Section II and III we will describe the experimental measurement and data/error analysis, respectively, while in Section IV the observed effect is explained theoretically based on spin-noise correlations that build up at low magnetic fields due to spin-exchange collisions.
\section{Measurement}
\begin{figure}
\includegraphics[width=8.5 cm]{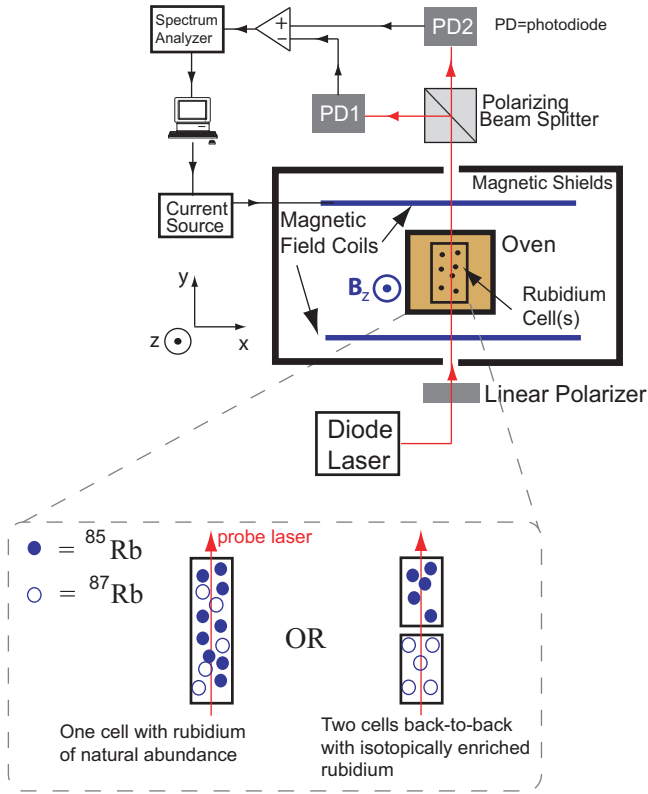}
\caption{(Color online). Experimental schematic of the spin noise measurement. For the actual measurement we used a 10 cm long cell with Rb of natural abundance, while for a consistency-check we used two 5 cm cells back-to-back, each having isotopically enriched Rb. The temperature was measured with a thermocouple placed at the oven's center, reading $112~^{\circ}{\rm C}$. The temperature inferred from the collisional line width of the spin noise resonance was  $100~^{\circ}{\rm C}$, and it was the corresponding Rb density that we used in the theoretical prediction. The laser power and detuning from the D2 line were 3.3 mW and 43 GHz, respectively, while the pressure-broadened optical linewidth at 100 torr of nitrogen is about 4 GHz. The magnetic field was set by a computer-controlled switch at either the desired value or a much larger value pushing spin noise out of the detector's bandwidth, enabling a fast subtraction of the background spectrum (no spin noise) from the spin-noise spectrum. The balanced polarimeter output was fed into a spectrum analyzer, and the spectra were averaged at the computer.}
\label{schem}
\end{figure}
The experimental scheme is shown in Fig.\ref{schem}, and is similar to previous studies of spin noise using a dispersive laser-atom interaction \cite{sn1,crooker,sn2,sn3,sn4}. An off-resonant laser illuminates a magnetically shielded rubidium vapor cell. A balanced polarimeter measures the Faraday rotation angle fluctuations of an initially linearly polarized and far-detuned laser. These fluctuations result from the fluctuating transverse spin, simultaneously precessing about a dc magnetic field transverse to the laser propagation direction.  As well known, at high laser detunings $\delta$ the Faraday rotation angle scales as $\theta\propto 1/\delta$ \cite{mabuchi}. Since the measured rotation signal is proportional to $\theta$ and to the laser power, both the laser wavelength and laser power were monitored and their fluctuations or drifts were less than 1\% and hence negligible.
\begin{figure}
\includegraphics[width=8.5 cm]{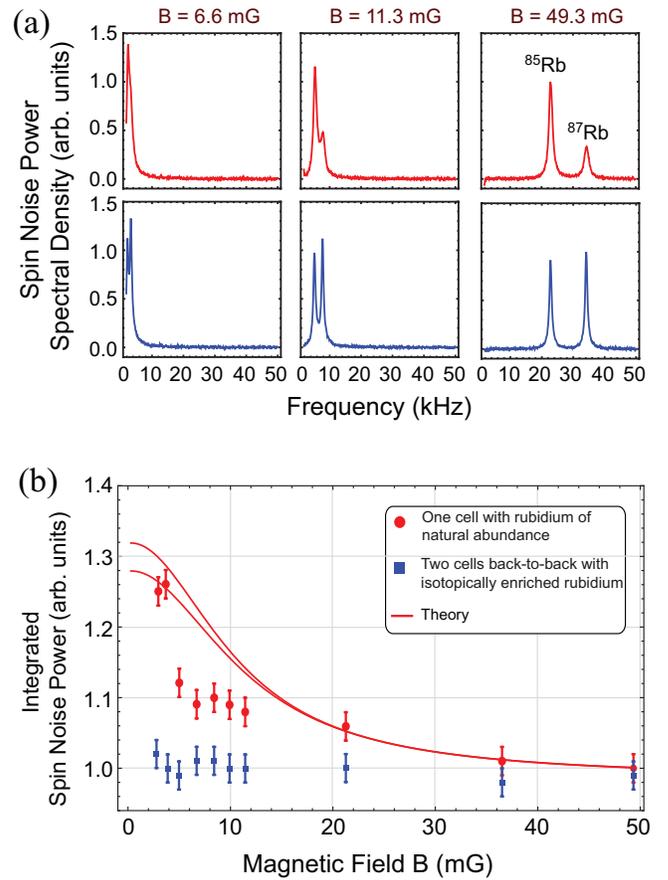}
\caption{(Color online). (a) Measured spin noise spectra for three different magnetic fields. Upper graphs are from a data set with the experiment cell (C1) containing Rb of natural isotopic abundance (ratio of peak heights about 3:1), and lower graphs are from
the two back-to-back cells (C2), each enriched with one of the two Rb isotopes (ratio of peak heights about 1:1).  (b) Integrated spin noise power (ISNP) for C1 (red circles) and C2 (blue squares). The former were normalized by their ISNP at $B=50~{\rm mG}$, while the latter were normalized by their average value. The red solid lines are the theoretical prediction $S(B)/S(50~{\rm mG})$ of Eq. \eqref{SB} with no free parameters, but with different (by 20\%) values of the magnetic gradient as input to the theory.}
\label{data}
\end{figure}
\begin{figure*}
\includegraphics[width=14 cm]{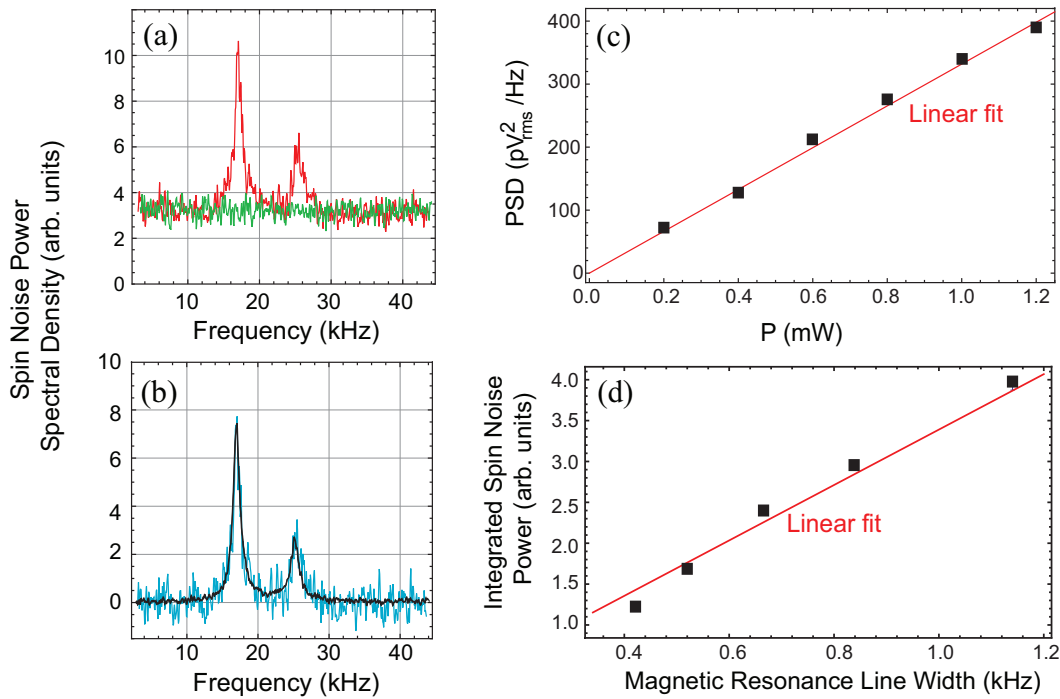}
\caption{(Color online). (a) Spin-noise spectrum and background. (b) The noisy blue line is the subtraction of the two spectra shown in (a) and constitutes a run, while the black line is the average of 50 runs and constitutes a set.  (c) The offset of the measured power spectra scales linearly with laser power, demonstrating a photon shot-noise limited measurement. (d) Total integrated spin-noise power at different temperatures. For the integral we used the $^{85}$Rb spin-noise resonance at high enough magnetic fields so that there is no overlap with the $^{87}$Rb resonance. The resonance line width is proportional to atom number probed by the laser. We have corrected for the other small contributions to line broadening and for the different average laser power in the cell at different temperatures. Spin-noise signals scale as $\sqrt{\rm atom~number}$, hence spin-noise power scales linearly with the line width.}
\label{analysis}
\end{figure*}
Typical spin-noise spectra at various magnetic fields are shown in Fig.\ref{data}a. They exhibit two peaks centered at the Larmor frequencies of $^{85}$Rb and $^{87}$Rb.  The spin-noise spectra at different magnetic fields are integrated, and the total spin-noise power is plotted in Fig.\ref{data}b. Interestingly, the total spin-noise power increases at low magnetic fields where the two magnetic resonance lines overlap. This noise increase is the experimental signature of spin-noise exchange. 

A consistency check was done to ensure the experiment's and analysis' ability to detect an actual change in spin-noise power. Instead of using a cell with rubidium of natural abundance, we performed the same measurement with two cells placed back-to-back, each enriched by one of the two rubidium isotopes. In this case there cannot be any inter-species spin-noise transfer, and the total spin-noise power is expected to be independent of the magnetic field, which is the case as shown in Fig.\ref{data}b. 
\section{Data and Error Analysis}
The integrated spin-noise power (ISNP) data of Fig.\ref{data}b were obtained in the following way. A time series of the polarimeter output was fed into a differential amplifier, the output of which was acquired by the spectrum analyzer (SA) having a measurement bandwidth of 50 kHz and a resolution bandwidth of 62.5 Hz. The corresponding measurement time is 16 ms. Sequentially, we measured the background by applying a large magnetic field to shift the spin noise way out of the 50 kHz bandwidth of the SA (Fig.\ref{analysis}a). The background spectrum was then subtracted from the spin noise spectrum.  A run consists of 100 averages of such subtracted spectra, and a data set consists of the average of 50 runs (Fig.\ref{analysis}b). 

The offset in the spectra of Fig.\ref{analysis}a is determined by photon shot noise (PSN), verified by the offset's linear dependence \cite{mueller} on laser power, depicted in Fig.\ref{analysis}c. As usual in noise-measurements, we also verified the linear scaling of the total spin-noise power with atom number, shown in Fig.\ref{analysis}d.
\begin{figure*}
\includegraphics[width=15 cm]{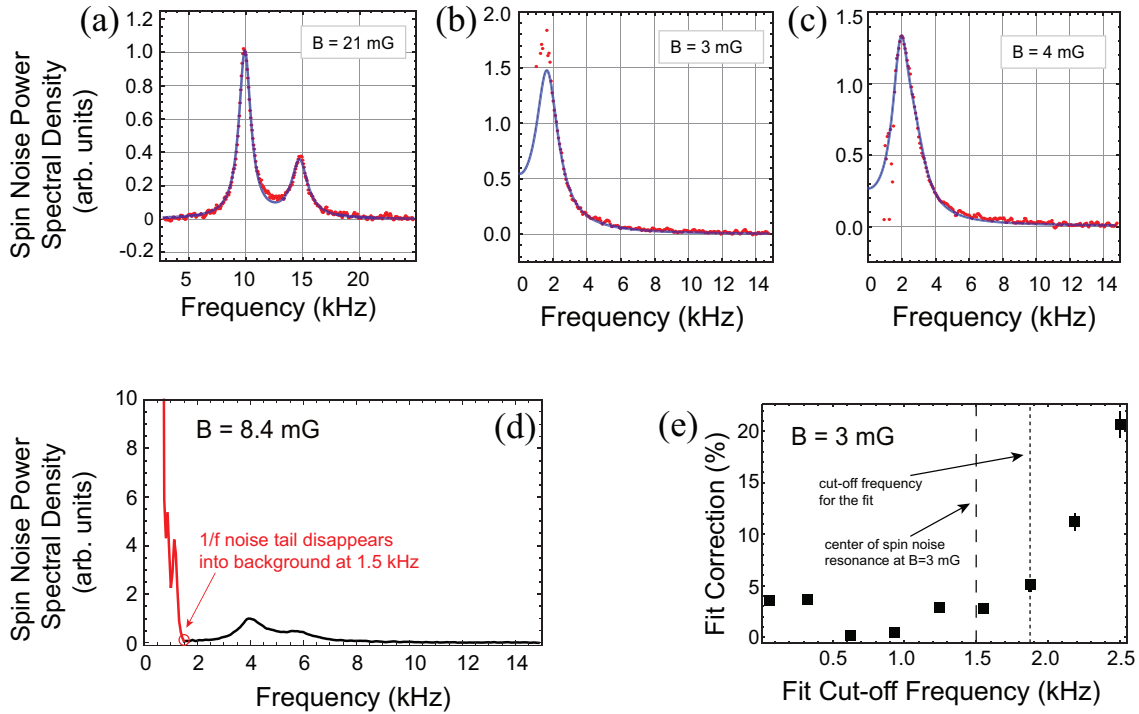}
\caption{(Color online) Spin noise spectrum with Lorentzian fit for (a) a high magnetic field, and (b,c) the lowest two magnetic fields. (d) The 1/f noise tail falls to the photon-shot-noise level at 1.5 kHz. (e) To avoid contamination of the $B$=3 mG data by the 1/f noise tail we start fitting the data from the fit cut-off frequency 1.9 kHz and on (short-dashed line). The long-dashed line at 1.5 kHz shows the line center frequency. To estimate the fit error due to the fit cut-off we generate numerical data with the same signal-to-noise ratio as the real data, fit them and compare with the known ISNP. For the $B$=4 mG data this error is negligible, since the peak of the resonance, being just higher than the cut-off of 1.9 kHz as shown in (c), is included in the fit.}
\label{analysis2}
\end{figure*}

For every magnetic field we measured three data sets both with the experiment cell and the two back-to-back cells. The ISNP in each set was calculated by fitting the spin-noise spectra with a Lorentzian lineshape, taking into account the negative frequency folding for the low-magnetic field spectra. The results of all sets were then averaged and presented in Fig.\ref{data}b. An example of spin noise data with the fit for a relatively high magnetic field is shown in Fig. \ref{analysis2}a, whereas Figs. \ref{analysis2}b and \ref{analysis2}c show the data and fit for the two lowest magnetic field points. To avoid contamination of the lowest magnetic field data ($B=3$ mG) by the 1/f noise tail we start fitting the data at 1.9 kHz, as the 1/f noise tail disappears into the PSN  background at 1.5 kHz (Fig. \ref{analysis2}d). This fit cut-off overestimates the true ISNP and needs to be corrected for. To estimate the correction we produce numerical data with the same signal-to-noise ratio as the real data and fit them starting from various cut-off frequencies. The ISNP of the numerical data is known, and the extracted fit correction is shown in Fig. \ref{analysis2}e.

For the higher magnetic fields we both fit the data with Lorentzians, and independently we numerically integrate the data to find ISNP. Both methods give perfectly consistent results. With the latter method we also estimate the ISNP error from the statistical distribution of the ISNPs of 50 runs. 
\section{Theoretical Explanation}
The theoretical explanation of the observed phenomenon follows by considering the detailed spin dynamics of a coupled spin ensemble. The three physical mechanisms driving single-species spin noise are (i)  damping of the transverse spin, (ii) transverse spin fluctuations and (iii) Larmor precession. Processes (i) and (ii) are both driven by atomic collisions, as also understood by the fluctuation-dissipation theorem \cite{sn2}. They involve (a) alkali-alkali spin exchange collisions and (b) alkali-alkali and alkali-buffer gas spin destruction collisions. Type-(b) collisions have a negligible cross-section compared to the spin-exchange cross section \cite{happer}
$\sigma_{se}=2\times 10^{-14}~{\rm cm}^{2}$, hence only type-(a) collisions will be considered. In the coupled double-species system there is an additional phenomenon: spin exchange collisions between different atoms. These are a sink of spin coherence for one atom and a source of spin polarization for the other. 
All of the above phenomena are compactly described by the coupled Bloch equations for the transverse spin polarizations $\mathbf{P}_{j}\equiv\mathbf{\hat{x}}\langle P_{j,x}\rangle+\mathbf{\hat{y}}\langle P_{j,y}\rangle$ of $^{85}$Rb ($j=1$) and $^{87}$Rb ($j=2$):
\beq
d\mathbf{P}_{1}=dt\left[\mathbf{P}_{1}\times\mbox{\boldmath$\omega$}_{1}-\gamma^{12}_{se}(\mathbf{P}_{1}-\mathbf{P}_{2})-\gamma_{1}\mathbf{P}_{1}\right]+d\mbox{\boldmath$\xi$}_{1}\label{dp1}
\eeq
\beq
d\mathbf{P}_{2}=dt\left[\mathbf{P}_{2}\times\mbox{\boldmath$\omega$}_{2}-\gamma^{21}_{se}(\mathbf{P}_{2}-\mathbf{P}_{1})-\gamma_{2}\mathbf{P}_{2}\right]+d\mbox{\boldmath$\xi$}_{2}\label{dp2}
\eeq
where $\mbox{\boldmath$\omega$}_{i}=\mathbf{\hat z}\omega_{i}=\mathbf{\hat z}g_{i}B$ are the Larmor frequencies of the two Rb isotopes in the magnetic field $\mathbf{B}=B\mathbf{\hat z}$. 
Similar equations, albeit for different binary mixtures and unrelated to spin-noise, have been used elsewhere \cite{romalis_he3,walker}.
\subsection{Relaxation rates and noise terms}
Spin exchange collisions transfer spin polarization from species $j$ to $i$ at a rate $\gamma^{ij}_{se}=\sigma_{se}\overline{v}n_{j}$, where $n_1$ and $n_2$ are the respective number densities and $\overline{v}$ the rms average relative velocity of the colliding atoms. The transverse spin relaxation rate of atom $j$ other than due to spin exchange with different-species atoms is given by $\gamma_{j}$ and consists of (i) spin-exchange with same-species atoms, $\gamma_{se}^{jj}=\sigma_{se}\overline{v}n_{j}$ and (ii) magnetic field gradient, $\gamma_{j,\nabla B}$. The total spin relaxation rate of atom $j$ will then be $\Gamma_{j}=\Gamma+\gamma_{j,\nabla B}$, where $\Gamma=\gamma_{se}^{11}+\gamma^{12}_{se}=\gamma_{se}^{22}+\gamma^{21}_{se}=\sigma_{se}\overline{v}(n_{1}+n_{2})$. From the fits of the noise peaks, and considering that  $\gamma_{2,\nabla B}=(g_2/g_1)^2\gamma_{1,\nabla B}$ \cite{catesGradient} it was found that for the 10 cm rubidium cell $\Gamma=2\pi\times 800~{\rm Hz}$, $\gamma_{1,\nabla B}=2\pi\times 300~{\rm Hz}$ and $\gamma_{2,\nabla B}=2\pi\times 700~{\rm Hz}$. For the two-cell measurement we found $\Gamma_1\approx\Gamma_2\approx \Gamma=2\pi\times 800~{\rm Hz}$, consistent with the fact that in this case the gradient relaxation is negligible since it scales with the 4$^{\rm th}$ power of cell dimension and the isotopic cells were 5 cm long each). There are two small additional relaxation sources common to both atoms: (i) the transit time through the probe laser, and (ii) probe laser power broadening. The former can be safely neglected. The latter is only 5\% of the total linewidth. Finally, $d\mbox{\boldmath$\xi$}_{j}$ ($j=1,2$) are independent Gaussian white noise processes with zero mean and variance $\Gamma dt/N_{j}$ \cite{noteS}, where $N_j$ is the total atom number of species-$j$ probed by the laser. 
\subsection{Integrated spin-noise power}
Introducing the 2-element column-vector $\mbox{\boldmath$\pi$}=(\pi_{1}~~\pi_{2})^{T}$, with $\pi_{j}=P_{j,x}+iP_{j,y}$, the Bloch equations (\ref{dp1}) and (\ref{dp2}) can be compactly written as 
\beq
d\mbox{\boldmath$\pi$}=-dt{\bf A}\cdot\mbox{\boldmath$\pi$}+\mbox{\boldmath$\Xi$}\cdot d\mathbf{W}\label{sigma}
\eeq
where the decay matrix is 
\beq
{\bf A}=\left(\begin{array}{ccc} 
\Gamma_{1}+i\omega_{1} & -\gamma^{12}_{se} \\
-\gamma^{21}_{se} & \Gamma_{2}+i\omega_{2}\\
\end{array}\right),
\eeq
and $\mbox{\boldmath$\Xi$}$ is the diagonal $2\times 2$ fluctuation matrix with $\Xi_{jj}=\sqrt{2\Gamma/N_{j}}$ and $j=1,2$. The noise vector $d\mathbf{W}=(dW_{1}~~dW_{2})^{T}$ describes two independent  complex Gaussian processes, $dW_{1}$ and  $dW_{2}$, having zero mean and variance $dt$ \cite{note}. 
The total spin $\sigma_y$ probed by the laser is the sum of the $y$-component of all rubidium atom spins inside the probe laser beam, $\sigma_y=\sum_{m=1}^{N}s_{m,y}$, which can be written as $\sigma_y=\mathfrak{Im}\{n_1\pi_1+n_2\pi_2\}$. 
The total spin-noise power $S(B)$ as a function of the magnetic field $B$ can be computed as 
\begin{align*}
S(B)&={1\over T}\int_{0}^{T}{dt\sigma_y^2(t)}={1\over {2T}}\int_{0}^{T}{dt|\sigma(t)|^2}\\&={1\over {2T}}\int_{0}^{T}{dt|n_1\pi_1(t)+n_2\pi_2(t)|^2}.
\end{align*}
Since the averaging time $T$ is much longer than the spin relaxation time, ergodicity of the Ornstein-Uhlenbeck process \mbox{\boldmath${\pi}$} ensures that the preceding long time average can be computed as an expectation under its equilibrium distribution. Now, \mbox{\boldmath${\pi}$} is a two-dimensional complex Gaussian process. Its equilibrium distribution has mean 0, while the covariance matrix $\Sigma$ with $\Sigma_{ij}=\EE\big[\pi_i\pi_j^*\big]$ for $i=1,2$ can be computed \big(cf \cite{gardiner} equation (4.4.51)\big) as the unique self-adjoint solution to the matrix equation  
\[
A\Sigma+\Sigma A^\dagger= \Xi\Xi^\dagger.
\]
Solving the system of linear equations we find
\begin{equation}
\Sigma_{ii}=\frac{\Gamma}{\Gamma_iN_i}\left(1+\frac{{\ga\gb}}{Q}\right)\quad\text{for }i\in\{1,2\}
\label{Sdiag}
\end{equation}
and
\begin{equation}
\Sigma_{12}=\Sigma_{21}^*=\frac{\Gamma\sqrt{\ga\gb}}{Q\sqrt{N_1N_2}} \left(1+i\frac{\Delta\omega}{\Gamma_1+\Gamma_2}\right)
\label{Soffdiag}
\end{equation}
\normalsize
where $\Delta\omega=\omega_2-\omega_1$ and 
\[
Q={\Gamma_1\Gamma_2\left[1+\left(\frac{\Delta\omega}{\Gamma_1+\Gamma_2}\right)^2\right]-\ga\gb}.
\]
Hence, $S(B)=\frac{1}{2}\EE\big[\ |n_1\pi_1+n_2\pi_2|^2\big]=\frac{1}{2}\mathbf{n}^T\Sigma\mathbf{n}$, where $\mathbf{n}^T=(n_1,n_2)$, and finally we get
\beq
{{S(B)}\over {S(\infty)}}={1\over {1-{{\gamma_{se}^{12}\gamma_{se}^{21}}\over {\Gamma_1\Gamma_2}}f(B)}}\Big[1+{{n_1\gamma_{se}^{12}+n_2\gamma_{se}^{21}}\over {n_1\Gamma_2+n_2\Gamma_1}}f(B)\Big],\label{SB}
\eeq
where 
\beq
{1\over {f(B)}}={1+{{4\gamma_{se}^{12}\gamma_{se}^{21}}\over{(\Gamma_1+\Gamma_2)^2}}\Big({B\over B_0}\Big)^2}. 
\eeq
Here we have defined $B_0^2=4\gamma_{se}^{12}\gamma_{se}^{21}/(g_2-g_1)^2$. For our experimental parameters $B_0\approx 3.5~{\rm mG}$. Equation \eqref{SB} leads to the theoretical prediction plotted in Fig. 2b with no free parameters.

For the ideal case of no magnetic gradient, $\gamma_{1,\nabla B}=\gamma_{2,\nabla B}=0$, the field $B_0$ signifies a transition from a high-field regime $B\gg B_{0}$ where the eigenvalues $\gamma=\Gamma+i(\omega_{1}+\omega_{2})/2\pm{{\omega_2-\omega_1}\over 2}\sqrt{B_0^2/B^2-1}$ of the decay matrix ${\bf A}$  describe two independent spin precessions at $\omega_1$ and $\omega_2$ and decaying at a rate $\Gamma$, to a low-field regime $B\ll B_{0}$ where the spin-exchange coupling forces the atoms to precess together at $(\omega_1+\omega_2)/2$, the precession having two decay rates ${\rm Re}\{\gamma\}=\Gamma\pm\sqrt{\gamma_{se}^{12}\gamma_{se}^{21}}$ \cite{note4,HapperTang,note5}. In this experiment the lowest field used is just about $B_0$ and this transition of the decay rates $\gamma$ is not observable. Further, in the absence of magnetic gradient the spin-noise power at zero field takes on the simple form 
\beq
{{S(0)}\over {S(\infty)}}={{r^2+4r+1}\over{r^2+r+1}}
\eeq
where $r\equiv n_1/n_2$. The excess spin-noise power is maximized for $r=1$, the maximum being 100\%, i.e. the spin-noise power is double at low fields relative to high fields.
\subsection{Spin-noise correlations} 
Towards explaining the observed effect we note that the off-diagonal elements of the covariance matrix $\Sigma$ carry information about the correlation of polarizations $\mathbf{P}_1$ and $\mathbf{P}_2$. It is $\mathbb{E}[\mathbf{P}_1\cdot\mathbf{P}_2]=\mathbb{E}[\mathfrak{Re}\{\pi_1\pi_2^*\}]=\mathfrak{Re}\{\Sigma_{12}\}$. We can thus compute the correlation coefficient 
\beq
\rho(B)\equiv{{\mathbb{E}[\mathbf{P}_1\cdot\mathbf{P}_2]}\over{\sqrt{\mathbb{E}[|\mathbf{P}_{1}|^2]\mathbb{E}[|\mathbf{P}_2|^2]}}}=\sqrt{{\gamma_{se}^{12}\gamma_{se}^{21}}\over {\Gamma_1\Gamma_2}}f(B)
\label{correl}
\eeq
Again, in the ideal case of no gradient relaxation it is $\rho(0)=\sqrt{r}/(1+r)$, which is also maximized for $r=1$ with the maximum being 1/2. Also, $\rho(B)\rightarrow 0$ when $B\gg B_{0}$. This leads to an intuitive explanation of the observed phenomenon as an exchange of spin-noise between two atomic species. 
In the rotating frame of atom $i$ the transverse spin of atom $j$ precesses at the frequency $\delta\omega=|\omega_2-\omega_1|$. If $\delta\omega\gg\Gamma$, in other words if the two spin noise resonances are far apart, the spin polarization of atom $j$ seen in the rotating frame of atom $i$ averages out to zero within the spin-exchange time of $1/\Gamma$. If, however, $\delta\omega\leq\Gamma$, then the noise polarization of atom $j$ transferred to $i$ adds up, to some extent coherently due to the non-zero $\rho(B)$, to the noise polarization of $i$. This is due to the strong polarization-noise correlations produced by spin-exchange. Hence the total spin-noise power is increased relative to the case where the two noise powers add just in quadrature for $\delta\omega\gg\Gamma$. 

To quantify the above discussion, let $\Pi_{i}$ be the total {\it power} of atom-$i$ polarization fluctuations. We can think of $\Pi_{i}=\Pi_{i}^{(0)}+\Pi_{ij}$ as consisting of two terms, the noise power $\Pi_{i}^{(0)}$ that we would observe if atoms-$i$ were alone, and the transfer of polarization noise from $j$ to $i$, described by the term $\Pi_{ij}$. Clearly, $\Pi_{i}^{(0)}=\Gamma/(\Gamma_in_{i})$. In view of (\ref{Sdiag}) we find that indeed $\Pi_{i}=\Pi_{i}^{(0)}+\Pi_{ij}$, with $\Pi_{ij}=\Pi_{j}[(\gamma_{se}^{ij})^{2}/\Gamma_{i}^2]f(B)$.
\begin{figure}
\includegraphics[width=7 cm]{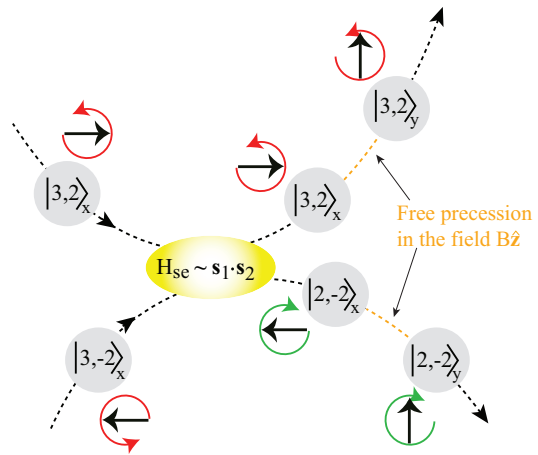}
\caption{(Color online). Example of a spin-exchange collision generating spin noise. The spin states $|F,m_{F}\rangle$ of the colliding atoms are written in the $x$-basis. Two $^{85}$Rb atoms in the $F=3$ state and opposite $m_F$ collide, and after the collision one jumps to the $F=2$ manifold. The total $m_F$ is conserved, however the spin in the $F=2$ state precesses in the opposite sense. Under the action of the magnetic field $B\mathbf{\hat{z}}$ the two atoms will momentarily generate a non-zero contribution to the spin-noise signal along the y-axis (balanced of course by the nuclear spins). Subsequent spin-exchange collisions will damp the transverse spin and so on and forth. }
\label{col}
\end{figure}
\subsection{Discussion}
For completeness we note the following. (i) In the two hyperfine levels of rubidium the spin precesses in opposite directions, corresponding to positive and negative frequencies. In the measured power spectrum both appear at the same positive frequency. (ii) Spin noise is genuine quantum noise produced by atomic collisions. The linear scaling of the total spin-noise power with atom number (Fig. \ref{analysis}d) does not by itself prove the previous assertion. Instead, the physics of spin-noise generation must be understood. Spin exchange collisions have two roles: they damp spin coherence and they generate noise coherence. As well known \cite{happer_walker}, atoms can jump from one hyperfine level to the other during a spin-conserving spin-exchange collision, thereby perturbing their coherent spin precession and leading to loss of spin coherence. The same mechanism can generate fluctuations of spin coherence as shown in the example of Fig. \ref{col}. In every collision there are a number of potential final states, the probability of which is determined by the quantum spin-dependent scattering of the atoms \cite{happer_book}, and hence spin-noise bears the fundamental quantum-mechanical unpredictability. (iii) There is an apparent disagreement between data and theory at intermediate-field data. The magnetic gradient was found to have a 20\% variation with magnetic field. In Fig. 2b we plot the theoretical prediction with a constant value for the gradient, but we haven shown how the theoretical prediction is affected by changing this constant value within its observed variation range. Either an unidentified systematic or a statistical outlier effect could be responsible for the aforementioned discrepancy. To demonstrate the spin-noise effect presented in this work without the added complication of magnetic gradients and with better statistics a short cell in the multi-pass arrangement of Romalis and co-workers \cite{multipass} would be most appropriate.
\section{Conclusions}
Concluding, we have experimentally demonstrated the interspecies transfer of spin noise through the spin-exchange coupling of two alkali vapors. This transfer, also seen as a positive correlation of the two-species polarization noise, manifests itself as a total noise power increase at low-magnetic fields, or to put it differently, as the decrease of the total spin-noise power at high fields where the spin-noise correlation vanishes.  Although we demonstrated the phenomenon using an unpolarized spin state, the same phenomenon would occur in the coherent spin state of a maximally polarized spin ensemble \cite{schori}, directly relevant to precision metrology applications. 
\acknowledgements
We acknowledge the anonymous Referees for their constructive criticism. A.T.D. acknowledges support by the European Union (European Social Fund ESF) and the Greek  Operational Program
"Education and Lifelong Learning" of the National Strategic Reference
Framework (NSRF) - Research Funding Program "Heracleitus II. Investing in
knowledge society through the European Social Fund". 
M.L. acknowledges support from NSRF Research Funding Programs Thales MIS377291 and Aristeia 68/1137-1082. I.K.K. acknowledges helpful discussions with Profs. M. Romalis and W. Happer and support from the European Union's Seventh Framework Program FP7-REGPOT-2012-2013-1 under grant agreement 316165.


\begin{thebibliography}{0}
\bibitem{noteS}
It is only $\Gamma$ that is responsible for spin fluctuations, as the gradient relaxation is not collisional relaxation and hence cannot generate fluctuations.
\bibitem{note}
A complex Gaussian process $dW$ with variance $dt$ is given by $dW=(dW_1+idW_2)/\sqrt{2}$, where $dW_1$ and $dW_2$ are independent real Gaussian 
processes with variance $dt$.
\bibitem{note4}
For this discussion we consider the ideal case where $\gamma_{1,\nabla B}=\gamma_{2,\nabla  B}=0$.
\bibitem{note5}
These eigenvalues are not to be used for too low a magnetic field, because when $\Gamma\omega_{1}\gg 1$ the effect of the suppression of spin-exchange relaxation \cite{HapperTang,magn1} will gradually dominate.
\end{thebibliography}
\end{document}